\documentclass[12pt]{article}
\usepackage{amsmath}
\usepackage{amsfonts}
\usepackage{graphicx}
\usepackage{color}
\newcounter{mnotecount}[section]

\renewcommand{\themnotecount}{\thesection.\arabic{mnotecount}}

\newcommand{\mnote}[1]
{\protect{\stepcounter{mnotecount}}$^{\mbox{\footnotesize
$
\bullet$\themnotecount}}$ \marginpar{
\raggedright\tiny\em
$\!\!\!\!\!\!\,\bullet$\themnotecount: #1} }

\def\be{\begin{equation}}
\def\ee{\end{equation}}
\def\bea{\begin{eqnarray}}
\def\eea{\end{eqnarray}}

\begin{document}
\title{Conformal Properties of an Evaporating Black Hole Model}
\author{Christian L\"ubbe\\Erwin Schr\"odinger Institute,\\ Boltzmanngasse 9,\\ A-1090 Vienna, Austria\\\\Paul Tod\\Mathematical Institute,\\ University of Oxford,\\  Oxford, OX1 3LB,UK}

\maketitle
\begin{abstract}
We use a new, conformally-invariant method of analysis to test
incomplete null geodesics approaching the singularity in a model of
an evaporating black hole for the possibility of extensions of the
conformal metric. In general, a local conformal extension is possible from
the future but not from the past.


\end{abstract}
\section{Introduction}
In \cite{RP}, Penrose developed his argument that, in the correct
theory of space-time, initial singularities and final singularities
would be different in character, with the Weyl tensor finite or even
zero at initial singularities but no such constraint on final
singularities. As a test case, he discussed the space-time
singularities associated with Hawking's picture of an evaporating
black hole (see also \cite{rp2}). In this picture, there is a final singularity formed in
the collapse which led to the black hole. At the endpoint of the
evaporation of this black hole, this singularity vanishes but not
before, at the last instant, being visible in the exterior and
therefore being briefly an initial singularity. Penrose's argument
then suggests that the singularity as seen from the past should be
different in character, with respect to conformal properties, from
the singularity as seen from the future.

To test this suggestion, we need a convincing model of the metric
for an evaporating black hole and we shall use the model of
Hiscock \cite{his1}, \cite{his2}. Essentially, this consists of
matching Vaidya metrics to the Schwarzschild metric of the black
hole and to flat space, with an ingoing flux of negative energy
density which reduces the Schwarzschild mass to zero, and an
outgoing flux of positive energy density which radiates the mass to
infinity, leaving flat space to the future of an outgoing null
hypersurface. Details will be given below, but note that the model
is spherically-symmetric.

Once we have the model, we test its conformal properties using a
method developed in \cite{lub}. Null geodesics are
conformally-invariant as point sets, in the sense of being unchanged
under conformal rescalings of the metric, and as a point set a null
geodesic is also a null conformal geodesic (see e.g. \cite{LT} for
the general definition of conformal geodesic). By regarding a null
metric geodesic, say $\gamma$, as a null conformal geodesic, one can
give a conformally-invariant definition of propagation of frames
along $\gamma$ and can then give conformally-invariant definitions
of boundedness of conformal curvature and its derivatives along
$\gamma$, namely by requiring components of the
(conformally-invariant) conformal curvature in these
conformally-invariant frames to be bounded. In general, it is more
natural to work with the calculus of tractors and tractor curvature
(see \cite{LT}). For our purposes in this article this technology
isn't needed, but we recall that the \emph{tractor curvature}, which
is the curvature of the tractor connection, has as nonzero
components the Weyl tensor $C_{abc}^{\;\;\;\;\;d}$ and the Cotton
tensor, which is the derivative $\nabla_{[a}P_{b]}^{\;\;d}$ of the
Schouten tensor (defined below). Now suppose that $\gamma$ is
incomplete as a metric null geodesic but that there is a conformal
rescaling of the metric which allows an extension of the space-time
so that $\gamma$ can be extended as a null geodesic of the rescaled
metric. Then necessarily the conformal curvature (and the tractor
curvature) will be bounded in the rescaled metric, so that the
conformally-invariant conditions of boundedness will hold. (It is a
recent result of one of us \cite{cl1} that these necessary
conditions are sufficient: given bounded derivatives up to order $k+1$  of the
tractor curvature, a $C^k$ local extension of the conformal
structure will exist.)

This method was used in \cite{lub} in a variety of particular cases:
on radial null and space-like geodesics in the Schwarzschild metric,
to find, as expected, that no conformal extension is possible
through the singularities at $r=0$ or through space-like infinity
$i_0$; on radial null geodesics of the \emph{conical spherical}
metric:
\[ds^2=dt^2-dr^2-a^2r^2(d\theta^2+\sin^2\theta
d\phi^2),\mbox{    }a\neq 1,\] (so-called because it has a
singularity at the origin due to a deficit in the solid angle,
analogous to the deficit in the plane angle at the familiar conical
singularity) to show that no conformal extension is possible through
the singularity at $r=0$; and along the matter flow-lines of the
Lemaitre-Tolman-Bondi dust cosmologies to find conditions for
conformal extendibility through the singularity.

In this article, we use the method to analyse the singularity which
is formed in the collapse and then evaporates with the black hole.
This singularity has, in the Penrose diagram  (see figure 1), an interior 
and a final point before it vanishes. Our findings may be summarised as follows:
\begin{itemize}
\item
along radial in or outgoing null geodesics meeting the singularity
at an interior point, the physical affine parameter distance to the singularity is
finite, and the Weyl spinor can never be made finite by rescaling;
no conformal extension is possible (note that this includes regions
of the Schwarzschild singularity, so in particular, as one would expect, no
conformal extension is possible through there);
\item
along radial ingoing null geodesics from the past to the final point
of the singularity, the physical affine parameter
distance is finite; the Weyl spinor vanishes by continuity; the
tractor curvature vanishes if one imposes Hiscock's condition (\cite{his1}, \cite{his2}) of
finite rates of particle creation; by imposing more conditions on
the mass function of the Vaidya interior metric, one may make more
derivatives of the tractor curvature zero (and if all derivatives up to order $k+1$ are bounded then there is a local $C^k$-extension of the conformal structure, \cite{cl1}); if one imposes the decay rate on the mass function which follows naively from the Hawking mass-loss formula, then the tractor curvature is singular (and no local extension of the conformal structure is possible);

\item
along radial outgoing null geodesics from the past to the final point 
of the singularity, the situation is similar to that with the conical spherical metric: there are choices of conformal factor which lead to a bounded Weyl spinor, but then the singularity is infinitely far off in the rescaled metric; all other choices lead to an unbounded Weyl tensor; either way no conformal extension is possible (it is also worth noting that the  affine parameter distance to the singularity in the physical, unrescaled metric is infinite in this case);
\item
along past-directed, radial ingoing null geodesics from the future to the final point
of the singularity 
the physical affine parameter
distance is finite, and of course the tractor curvature and its derivatives are zero; local
conformal extension is possible.
\end{itemize}
In summary, the conclusions accord with Penrose's hypothesis: a conformal extension is possible from the future but not from the past, with the exception that local extensions of low differentiability may be possible along ingoing null geodesics to the final point, given extra conditions on the rate at which the mass-function vanishes. A global extension from the past, through the whole singularity, is never possible.

\medskip

The plan of the article is as follows. We begin in the next section
by giving more details of the evaporating black hole (EBH) models,
introducing a Newman-Penrose null tetrad and calculating what we
need from the spin-coefficient formalism (for which see e.g.
\cite{pr} or \cite{st}). In section 3, we give the details of the
method of testing for the possibility of a conformal extension, in
terms of null conformal geodesics and the conformally-invariant
propagation of spinors along them. In section 4, we collect the
results of applying this test along radial in and outgoing null
geodesics.

\section{The model}
As noted above, Hiscock's evaporating black hole (EBH) models
\cite{his1}, \cite{his2} are constructed by matching Vaidya metrics
to the Schwarzschild metric of the black hole and to flat space.
Start with a Schwarzschild metric, formed in collapse, and given to
the future of a constant $v$ surface ${\cal{S}}$, say $v=v_S$, in
the usual coordinates (so $v=t+r+2m\log(r-2m)$ where $m$ is the mass
of the initial Schwarzschild metric). The details of the collapse are
assumed to lie to the past of ${\cal{S}}$. Pick a sphere
$S=(u=u_1,v=v_1)$ on ${\cal{S}'}=\{v=v_1>v_S\}$ and a spherically-symmetric
time-like surface $\Sigma$ with past boundary at $S$. In the
space-time region to the future of  ${\cal{S}'}$ bounded by $\Sigma$, take
the metric to be the ingoing Vaidya metric (\ref{v1}) below, with mass
$N(v)$, where $N(v_1)=m$ and $N$ decreases to zero at $v_0$ (so $N(v_0)=0$) as negative mass
flows into the hole. In the space-time region to the future of ${\cal{S}''}=\{u=u_1\}$ bounded by $\Sigma$, take the metric to be the
outgoing Vaidya metric (\ref{v2}) below, with mass $M(u)$, where $M(u_1)=m$
and $M$ decreases to zero at $u_0$ (so $M(u_0)=0$) as positive mass flows out to future-null-infinity ${\mathcal{I}}^+$. $\Sigma$ has future boundary at the 2-sphere $S''=(u=u_0,v=v_0)$. In
the union of the regions to the past of ${\cal{S}'}$ and to the past
of ${\cal{S}''}$, retain the Schwarzschild metric.

\begin{figure}
\begin{center}
{\input{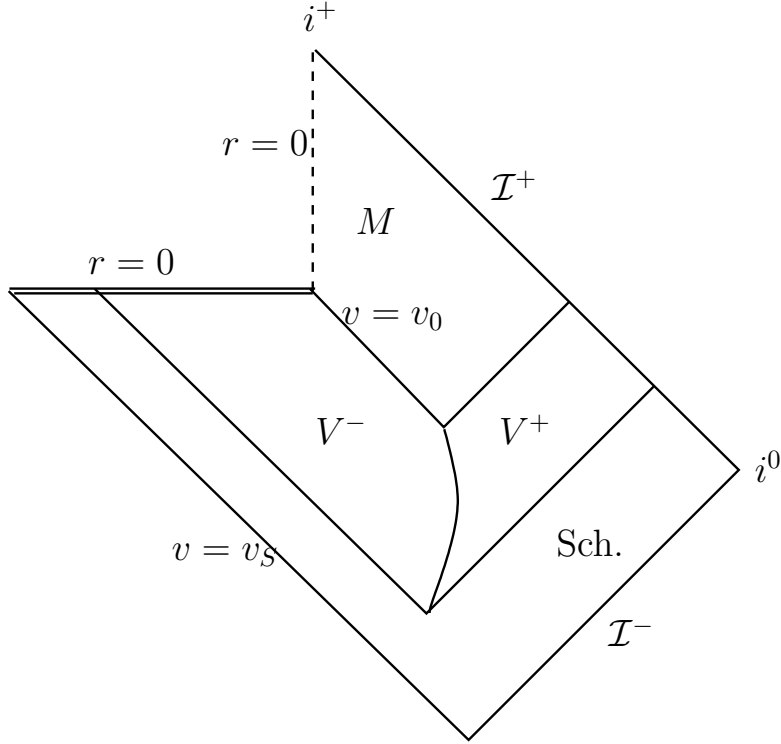}}

\caption{Penrose diagram of the model EBH space-time; $M$, $V^+$, $V^-$ and Sch. are regions of Minkowski space, outgoing Vaidya, ingoing Vaidya and Schwarzschild respectively; $V^+$ meets $V^-$ along the time-like hypersurface $\Sigma$, other matchings are along null hypersurfaces ($V^-$ to Sch. along $v=v_1$, $V^+$ to Sch. along $u=u_1$, $V^+$ to $M$ along $u=u_0$); the singularity at $r=0$ (double line) becomes the origin of coordinates in $M$ (dashed line) when the mass has dropped to zero (at $v=v_0$).}
\end{center}
\end{figure}

There will necessarily be distributional curvature along $\Sigma$,
and the choices of $N(v)$, $M(u)$, this distributional curvature and
the history of $\Sigma$ are tied together by the Einstein equations.
However, for our purposes, we don't need the details of this as long
as there is a sphere $(u=u_0,v=v_0)$ on $\Sigma$ as a future
boundary at which $N(v_0)=0=M(u_0)$. Assume so, and then match to
flat space across $v=v_0$ and across $u=u_0$. The Penrose
diagram for this metric is given in figure 1.

Note in
particular that $\Sigma$ must stop before the singularity is
reached. This is because at the singularity $r$ vanishes, while in
the outgoing Vaidya metric the singularity at $r=0$ is in the past,
corresponding to the past singularity of Schwarzschild, not the
future one.

Our interest is in the future singularity of the Schwarzschild
metric at $r=0$ (the past one is in a part of Schwarzschild not
relevant in collapse). Points of the future one with $v\leq v_1$ are
just as in Schwarzschild, while points with $v_1\leq v<v_0$ are as
in Vaidya. Our main interest is in points with $v=v_0$, and the
approach to them along different radial null directions. When
approached from the past like this, the last part of the approach is
solely through the part of the metric which is ingoing Vaidya;
approach from the future is through flat space; therefore we don't
need to calculate with the part of the metric which is outgoing
Vaidya, and so the details of $\Sigma$ don't affect our calculation.

The ingoing Vaidya metric is
\be \label{v1}
ds^2=\left(1-\frac{2N(v)}{r}\right)dv^2-2dvdr-r^2(d\theta^2+\sin^2\theta
d\phi^2)
\ee
and, although we shan't need it, the outgoing Vaidya metric is
\be \label{v2}
ds^2=\left(1-\frac{2M(u)}{r}\right)du^2+2dudr-r^2(d\theta^2+\sin^2\theta
d\phi^2) \ee
We choose the following Newman-Penrose (NP) tetrad for the ingoing
Vaidya metric:
\be \label{np1}
D=\partial_v+\frac12\left(1-\frac{2N(v)}{r}\right)\partial_r\,,\,\Delta=-\partial_r\,,
\ee
%
\[\delta=\frac{1}{r\sqrt{2}}\left(\partial_\theta+\frac{i}{\sin\theta}\partial_\phi\right),\]
%
so that $\ell^a$ is outgoing into the future and $n^a$ is ingoing.
Now calculate, using the NP formalism
\be\label{np3}
D\ell^a=2\epsilon\ell^a\,,\,Dn^a=-2\epsilon n^a\,,
\ee
and
\be \label{np4} \Delta\ell^a=0\,,\,\Delta n^a=0\,,
\ee
where
\be \label{np33}
\epsilon=\frac{N}{2r^2}.
\ee
 Note also that 
\be
\label{ro1}
\rho=-\frac{Dr}{r}=-\frac{1}{2r}\left(1-\frac{2N(v)}{r}\right),
\ee
so that the surface $r=2N(v)$ consists of marginally outer-trapped
surfaces of constant $r$, interpolating between the horizon of the
initial Schwarzschild solution at $r=2m$ on $v=v_S$ and the origin of flat
space at $v=v_0$. This surface is a dynamical horizon in the
language of \cite{ak}, but is diminishing in size as it has a flux
of negative energy across it.

The only nonzero curvature components are
\be\label{np5}
\Phi_{00}=\frac{N'}{r^2}\,,\,\Psi_2=-\frac{N}{r^3}.
\ee
so that
\be\label{cur1}
\Phi_{ab}=\Phi_{00}n_an_b\,,\,\Psi_{ABCD}=6\Psi_2o_{(A}o_B\iota_C\iota_{D)},
\ee
in terms of the spinor dyad underlying the chosen NP tetrad.

Hiscock \cite{his1}, \cite{his2} proposes imposing
the condition $N'(v_0)=0$, as this is necessary and sufficient to
give a finite rate of particle production at $v=v_0$. An alternative would be to take
\be\label{ha1} N(v)\sim(v_0-v)^{1/3},\ee
 which is the dependence obtained if one assumes that the Hawking expression for the power radiated by an evaporating black hole holds up to the end. In this case,  the Ricci tensor is singular at $v_0$. We'll return
to this point below, but note here that, away from $r=0$, Hiscock's condition has the
effect of making the curvature continuous.

\section{Conformal geodesics and the method of analysis}
A null conformal geodesic (see e.g. \cite{LT}) is a curve $\gamma$
with tangent vector $v^a$ and a covector $b_a$ defined along
$\gamma$ satisfying the coupled system
\bea
\nabla_vv^a+2(b_cv^c)v^a&=&0\label{cg1}\\
\nabla_vb_a-(b_cv^c)b_a+\frac12(b_cb^c)v_a&=&P_{ab}v^b\label{cg2}
\eea
where  $P_{ab}=\Phi_{ab}-\Lambda
g_{ab}=-\frac12R_{ab}+\frac{1}{12}Rg_{ab}$ is the Rho or
Schouten tensor (here we use the conventions of \cite{pr} since we use the spin-coefficient formalism; other authors may have permutations of the signs here). By (\ref{cg1}), a null conformal geodesic is a null
metric geodesic with a particular scaling, dictated by a chosen
solution of (\ref{cg2}).

Define a propagation law along conformal geodesics by
\be\label{cg19}\nabla_ve^a+(b_cv^c)e^a+(b_ce^c)v^a-(e_cv^c)b^a=0,\ee
which as we shall see has a good transformation under conformal rescaling, then in particular
$v^a$ is carried along by this rule. For spinors the propagation (\ref{cg19}) reduces to
\be\label{cg3}
\nabla_v\alpha^A=-v^{AA'}b_{A'C}\alpha^C
\ee
Under conformal rescaling
\be\label{res1}
\tilde{g}_{ab}=\Omega^2g_{ab}
\ee
we claim that solutions of (\ref{cg1})-(\ref{cg19}) transform according to
\[\tilde{v}^a=v^a\,,\,\tilde{b}_a=b_a-\Upsilon_a\,,\,\tilde{e}^a=e^a,\]
where, as usual, $\Upsilon_a=\Omega^{-1}\nabla_a\Omega$, and then
(\ref{cg3}) is invariant with $\tilde\alpha^A=\alpha^A$. Thus
(\ref{cg3}) provides a conformally-invariant propagation of, say, a
suitably normalised spinor dyad along $\gamma$, from which we
construct a conformally-invariant test for boundedness of Weyl
spinor components.

For the normalisation, note that, given two solutions $\alpha_1^A$
and $\alpha_2^A$ of (\ref{cg3})
\[\nabla_v(\epsilon_{AB}\alpha_1^A\alpha_2^B)  = -(b_cv^c)\epsilon_{AB}\alpha_1^A\alpha_2^B\]
so that, if $\Omega$ is a conformal factor with
\be\label{cg6}
v^a\tilde{b}_a:=v^a(b_a-\Upsilon_a)=0
\ee
then
\[\tilde\epsilon_{AB}\alpha_1^A\alpha_2^B:=\Omega\epsilon_{AB}\alpha_1^A\alpha_2^B\]
is constant along the conformal geodesic. This rescaling of $\epsilon_{AB}$ corresponds to the
rescaling (\ref{res1}) of the metric and then (\ref{cg1}) becomes
\[\tilde\nabla_vv^a=0,\]
where $\tilde\nabla$ is the metric covariant derivative for
$\tilde{g}$, so that $v^a$ is tangent to affinely-parametrised null
metric geodesics for $\tilde{g}$. It is then easy to see that
$u^a=\Omega^2v^a$ is affinely-parametrised for $g$. If $s$ and
$\tilde{s}$ are affine parameters for $u$ and $v$ respectively then
\be\label{affpar}
1=\nabla_v\tilde{s}=\Omega^{-2}\nabla_u\tilde{s}=\Omega^{-2}\frac{d\tilde{s}}{ds}.
\ee
Note that $\tilde{s}$ is a projective parameter along $\gamma$ in the usual terminology of conformal geodesics \cite{LT}, so that different choices of $v^a$ and $b_a$ subject to (\ref{cg1}) and (\ref{cg2}) lead to M\"obius transformations in $\tilde{s}$. Equation (\ref{cg6}) can be rewritten as
\[\Omega^{-1}\nabla_v\Omega=b_cv^c\]
and then the contraction of (\ref{cg2}) with $v^a$
gives
\be
\label{cg7}
\nabla_v\nabla_v\Omega^{-1}=\Omega^{-1} (P_{ab}v^av^b).
\ee
The method is now as follows: an affinely-parametrised null geodesic
$\gamma$ of the space-time metric, with tangent $u^a$, gives rise to
a null conformal geodesic by solving (\ref{cg2}) along $\gamma$ for
$b_a$, with $v^a$ replaced by $u^a$, and then adjusting the scaling
by solving (\ref{cg1}) for $v^a$; one can use (\ref{affpar}) to see
if the null geodesic is incomplete for $\tilde{s}$; if it is, one
can use a spinor dyad solving (\ref{cg3}) to test the curvature
components for boundedness on $\gamma$. If $\gamma$ is incomplete
for $\tilde{s}$ but with bounded curvature for $\tilde{g}$, then we
expect to be able to extend the unphysical metric in a neighbourhood
of $\gamma$ including a final segment \cite{LT}, \cite{cl1}. This
method therefore tests for extendibility of the conformal metric.

We shall apply this method to ingoing and outgoing radial null
geodesics in the ingoing Vaidya metric which encounter the
singularity at $r=0$, to test whether the conformal metric can ever
be extended through this singularity.

\section{The analysis}
\subsection{In-going null geodesics}
First we consider ingoing null geodesics, that is null geodesics
parallel to $n^a$. This geodesic vector field is
affinely-parametrised for the physical metric (by (\ref{np4}) so we
take $u^a=n^a$. Note that the coordinate $r$ is an affine parameter,
so that the singularity is at a finite distance in the physical
affine parameter, and the coordinate $v$ is constant along such a
$\gamma$. We distinguish $v=v_2<v_0$, an ingoing null geodesic
meeting the singularity in its interior, from $v=v_0$, an ingoing
null geodesic meeting the `end' of the singularity. If $v_2<v_1$,
then we are testing the Schwarzschild metric for extendibility,
while $v_1<v_2\leq v_0$ relates to the Vaidya metric.

 Expand $b_a$ in
the NP tetrad as
\[b_a=X\ell_a+Yn_a-\overline{Z}m_a-Z\overline{m}_a\]
and expand (\ref{cg2}) using (\ref{np4}) and (\ref{np5}) to obtain the system
\begin{eqnarray*}
\Delta X&=&X^2\\
\Delta Y&=&Z\overline{Z}\\
\Delta Z&=&XZ.
\end{eqnarray*}
We won't need $Y$. Note that, from (\ref{cg6}),
\[\Omega^{-1}\Delta\Omega=b_cu^c=X\]
from which we obtain $\Omega$, given $X$.

This system of equations is homogeneous (since, by (\ref{cur1}),
$P_{ab}n^b=0$) so that the solutions are as in vacuum . For $X$ and
$Z$, use (\ref{np1}) to find two cases:
\begin{enumerate}
\item
$X=0\,,\,Z=Z_0$ for constant $Z_0$, so $\Omega=1$ w.l.o.g.; or
\item
$X=(r+r_0)^{-1}\,,\,Z=Z_0(r+r_0)^{-1}$ for constants $r_0$ and $Z_0$, so $\Omega=(r+r_0)^{-1}$.
\end{enumerate}
Now for the spinor propagation (\ref{cg3}), expand
\[\alpha^A=\zeta o^A+\eta \iota^A\]
to find
\begin{eqnarray*}
\Delta\zeta&=&0\\
\Delta\eta&=&-X\eta-Z\zeta.
\end{eqnarray*}
A basis of solutions in the two cases is given by
\begin{enumerate}
\item
$O^A=o^A+(\eta_0+rZ_0)\iota^A\,,\,I^A=\iota^A$
\item
$O^A=o^A+(\eta_0(r+r_0)-Z_0)\iota^A\,,\, I^A=(r+r_0)\iota^A.$
\end{enumerate}
We wish to test the components of the Weyl spinor in these bases.
From (\ref{cur1}) in case 1 we find
\[\psi_2=\Psi_2=-\frac{N}{r^3}\]
which clearly diverges on the approach to the singularity at $r=0$. In case 2, we have
\[\psi_2=(r+r_0)^2\Psi_2\]
and, for any choice of $r_0$, this still diverges if $v=v_2<v_0$,
(which, by the remark above, includes the Schwarzschild singularity)
but it will be zero on $v=v_0$. To go further with the case $v=v_0$,
we calculate the other part of the tractor curvature, which is the
Cotton tensor $\nabla_{[a}P_{b]c}$ and which will be conformally
invariant at points where the Weyl tensor vanishes. For this Vaidya
metric, it is given by
\[\nabla_{[a}P_{b]c}=\nabla_{[a}\Phi_{b]c}=N'\nabla_{[a}\left(\frac{1}{r^2}n_{b]}n_c\right),\]
with the aid of (\ref{np5}) and (\ref{cur1}), which is
\[\frac{N'}{r^3}\left(3\ell_{[a}n_{b]}n_c+n_{[a}g_{b]c}\right)\]
The component of this along
$I^A\overline{I}^{A'}O^BO^C\overline{O}^{B'}\overline{O}^{C'}$ is
$N'(v_0)/r^3$ in case 1 and $N'(v_0)\{(r+r_0)^2/r^3\}$ in case 2,
which diverges in both cases unless $N'(v_0)=0$. The vanishing of
this, which is necessary for finiteness of the tractor curvature, is
Hiscock's condition \cite{his1}, \cite{his2} for
finite rates of particle creation at the end of the black-hole
evaporation, so that there is a physical motivation for imposing it.
If we do, then the whole of the tractor curvature is zero on this
null geodesic (If we used (\ref{ha1}) instead, then the tractor curvature is singular and no extension is possible.)

To test for the existence of a conformal extension, we need to consider higher derivatives of the tractor curvature.  
We next calculate the Bach tensor:
\[
B_{ab}=2\nabla_{A'}^{\;\;C}\nabla_{B'}^{\;\;D}\Psi_{ABCD}+2\Psi_{ABCD}\Phi_{A'B'}^{\;\;\;\;\;\;\;\;CD}.
\]
Recall that, under (\ref{res1}), $\tilde{B}_{ab}=\Omega^{-2}B_{ab}$.
When $v=v_0$, if $N=N'=0$, the only nonzero term in $B_{ab}$ will be
\[-\frac{12}{r^3}o_{(A}o_B\iota_C\iota_{D)}\nabla_{A'}^{\;\;C}\nabla_{B'}^{\;\;D}N(v)\]
\[=-\frac{2}{r^3}n_an_bN'',\]
using (\ref{np1}). Now components of $\tilde{B}_{ab}$ in frames of
either case diverge unless $N''=0$, so there is no conformal
extension possible unless $N''=0$. Inductively, if $N^{(k)}(v_0)=0$
for $0\leq k\leq n-1$ then the derivatives
\[\nabla_{a_1}\cdots\nabla_{a_k}C_{bcd}^{\;\;\;\;\;\;e}\]
will be zero for $0\leq k\leq n-1$, and the case  $k=n$ will be a
conformally-invariant tensor, whose only nonzero term will be
proportional to $N^{(n)}r^{-3}$; so the $(n-1)$-th derivative of
conformal (or tractor) curvature will be bounded if and only if
$N^{(n)}(v_0)=0$, and then a local $C^{n-2}$-extension of the conformal structure will exist, by \cite{cl1}.

\subsection{Outgoing null geodesics}
Next we consider outgoing null geodesics, in the direction of
$\ell^a$. This vector field is not affinely-parametrised (see
(\ref{np3})), so we introduce $u^a=\Theta^2\ell^a$ which is. By
(\ref{np3}), the condition on $\Theta$ is
\[0=\nabla_uu^a=\Theta^2D(\Theta^2\ell^a),\]
so that
\be\label{the1} D\Theta=-\epsilon\Theta.
\ee
Again, we expand $b_a$ in the NP tetrad, to find for (\ref{cg2}) the
system 
\bea
DX+2\epsilon X-Z\overline{Z}&=&0\nonumber\\
DY-2\epsilon Y-Y^2&=&\frac{N'}{r^2}\label{b21}\\
DZ-YZ&=&0.\label{b22}
\eea
and for the spinor propagation
\bea
D\zeta+\epsilon\zeta&=&-Y\zeta-\overline{Z}\eta\label{z21}\\
D\eta-\epsilon\eta&=&0.\label{z22}
\eea
For the conformal factor
\be\label{om1} \Omega^{-1}D\Omega=b_c\ell^c=Y,
\ee
Use (\ref{om1}) to solve (\ref{b22}) and find $Z=\Omega Z_0$,  then
(\ref{b21}) becomes a linear equation for $\Omega^{-1}$ :
\be\label{om2}
D^2\Omega^{-1}+2\Theta^{-1}D\Theta
D\Omega^{-1}+\Omega^{-1}\frac{N'}{r^2}=0
\ee
which is equivalent to (\ref{cg7}). Using (\ref{ro1}), one solution is found to be $\Omega=r^{-1}$,
corresponding to $Y=\rho$ . The general solution can therefore be written as
\be\label{om3}\Omega^{-1}=rF\ee
whereupon (\ref{om2}) can be integrated to give
\be\label{om4}DF=Ar^{-2}\Theta^{-2}\ee
for constant $A$.

From (\ref{the1}) we may solve (\ref{z22}) by
\[\eta=\eta_0\Theta^{-1}.\]
Now use (\ref{the1}) and (\ref{om1}) to convert (\ref{z21}) to
\[D(\Theta^{-1}\Omega\zeta)=-\overline{Z}_0\eta_0\Omega^2\Theta^{-2}.\]

A normalised spinor dyad is
\begin{eqnarray*}
O^A&=&\Theta\Omega^{-1}o^A\\
I^A&=&Wo^A+\Theta^{-1}\iota^A
\end{eqnarray*}
where
\[D(\Theta^{-1}\Omega W)=-\overline{Z}_0\Omega^2\Theta^{-2}.\]
In this dyad, the nonzero Weyl curvature components are
\bea
\psi_2&=&\Omega^{-2}\Psi_2 , \label{si1}\\
\psi_3&=&3\Theta^{-1}W\Omega^{-1}\Psi_2, \label{si2}\\
\psi_4&=&6W^2\Theta^{-2}\Psi_2. \label{si3}\eea
We need the behaviour of these along $\gamma$, where
\[\frac{dr}{dv}=\frac12-\frac{N(v)}{r},\]
which we need to solve for $r(v)$. We distinguish two cases:
\begin{itemize}
\item
$v\rightarrow v_2<v_0$: this includes the case of approach to the
Schwarzschild singularity.

Then
\[r=(2N(v_2))^{1/2}(v_2-v)^{1/2}(1+O(v_2-v)),\]
so
\[\Theta=(v_2-v)^{1/4}(1+O(v_2-v)).\]
From (\ref{om3})-(\ref{om4}), $\Omega^{-1}$ is therefore
asymptotically $O(1)$ or $O((v_2-v)^{1/2})$ as $v\rightarrow v_2$.
From (\ref{si1}) and (\ref{np5}), we can't make $\psi_2$ bounded. 

From the definition of $\Theta$, the physical affine parameter $s$ is obtained by solving
\[\frac{ds}{dv}=Ds=\Theta^{-2},\]
so that
\be\label{affp}\int ds=\int\Theta^{-2}dv.\ee
Now the distance to the singularity in the physical affine parameter is finite.
\item $v\rightarrow
v_0$:
 now we need to make an assumption about the behaviour of $N(v)$ subject to
 $N(v_0)=0$; we shall suppose that
 \[N(v)=n(v_0-v)^k\]
with $k>1$,  so that Hiscock's condition $N'(v_0)=0$ holds, then along $\gamma$ we calculate
\[r=2n(v_0-v)^k\left(1-4nk(v_0-v)^{k-1}+O((v_0-v)^{2(k-1)})\right).\]
Solve (\ref{the1}), using (\ref{np33}), to find
\be
\label{the2}
\Theta=(v_0-v)^k\exp\left(-\frac{1}{8n(k-1)(v_0-v)^{k-1}}\right)(1+O((v_0-v)^{k-1})).
\ee
From (\ref{affp}) and (\ref{the2}), the integral for $s$ diverges  as $v\rightarrow v_0$,
so that this singularity is infinitely far away along $\gamma$, as
measured in the physical affine parameter. Note that this is
different from the previous case of radial null geodesics ingoing
towards this same singularity.

From (\ref{om3}), (\ref{si1}) and (\ref{np5}):
\[\psi_2=-\frac{F^2N}{r}=-\frac{F^2}{2}(1+O((v_0-v)^{k-1})),\]
while, from (\ref{om4})
\be\label{F3}\frac{dF}{dv}=\frac{A'}{(v_0-v)^{4k}}\exp\left(\frac{1}{4n(k-1)(v_0-v)^{k-1}}\right)(1+O((v_0-v)^{k-1}))\ee
with $A'=A(4n^2)^{-1}$. 

We distinguish two cases. If $A=0$ then $F$ is constant and $\psi_2$ is finite. By choosing $Z_0=0=W$, we ensure that the whole Weyl spinor is bounded (from (\ref{si2}), (\ref{si3})). From (\ref{affpar}) however, choosing $F=1$ we find that
\[\frac{d\tilde{s}}{dv}=r^{-2}\Theta^{-2},\]
and the right-hand-side of this is the same as in (\ref{F3}), up to a nonzero factor. This equation has no solution which is bounded as $v\rightarrow v_0$, so that the singularity is infinitely far off in the rescaled, unphysical metric - no conformal extension is possible.

Now if $A\neq 0$ then we have to solve (\ref{F3}) for $F$ and again this has no solution 
which is bounded as $v\rightarrow v_0$, so that $\psi_2$ cannot be
made finite - in this case the distance to the singularity in the unphysical affine parameter is finite since
\[\frac{d\tilde{s}}{dv}=\Omega^2\Theta^{-2}=r^{-2}F^{-2}\Theta^{-2}=A^{-1}F^{-2}\frac{dF}{dv},\]
so that
\[\tilde{s}=c_1+\frac{c_2}{F},\]
which is bounded, but the Weyl curvature is singular so that no conformal extension is possible this way either. With Hiscock's condition, no local conformal extension is possible around the outgoing null geodesic to $v=v_1$.

It is straightforward to repeat the last calculation with (\ref{ha1}) instead. Now the physical affine parameter distance to the singularity is finite, but no choices lead to finite Weyl spinor, so that no local conformal extension is possible around this null geodesic.

\end{itemize}

\end{document}